\documentclass[aps,prb,reprint, floatfix,preprintnumbers,showpacs,amsmath,amssymb]{revtex4-1}
\usepackage{graphicx}
\usepackage{hyperref}

\begin{document}

\title{Incommensurate spin Luttinger liquid phase in a model for the spin-Peierls materials TiOBr and TiOCl}
\date{\today}

\author{D. Mastrogiuseppe}
\author{C. Gazza}
\author{A. Dobry}
\affiliation{Facultad de Ciencias Exactas Ingenier{\'\i}a y Agrimensura, Universidad Nacional de Rosario and Instituto de
F\'{\i}sica Rosario, Bv. 27 de Febrero 210 bis, 2000 Rosario,
Argentina.}

\begin{abstract}

In the present work we aim to characterize the lattice configurations and the magnetic behavior in the incommensurate phase of 
spin-Peierls systems. This phase emerges when 
the magnetic exchange interaction is coupled to the distortions of an underlying triangular lattice and has its experimental realization in the 
quasi-one dimensional compound family TiOX (X = Cl, Br). With a simple model of spin-$1/2$ chains inserted in a planar triangular geometry
which couples them elastically, we are able to obtain the uniform-incommensurate and incommensurate-dimerized phase transitions seen in
these compounds. Moreover, we follow the evolution of the wave-vector of the distortions with temperature inside the
incommensurate phase. Finally, we predict gapless spin excitations for the intermediate phase of TiOX compounds along with incommensurate spin-spin correlations,
offering to experimentalists a candidate to observe an exotic Luttinger liquid-like behavior.

\end{abstract}
\maketitle

\section{Introduction}
The recent discovery of a new class of inorganic spin-Peierls (SP) materials, TiOX (X = Cl, Br), has made 
reemerge the interest on this magneto-elastic transition. Unlike the previously known SP
materials, these compounds present a novel intermediate temperature incommensurate phase between the usual
uniform and dimerized ones. Other peculiar properties are the presence of a pseudo spin gap up to $T^*\sim 135K$ \cite{imai_chou} for TiOCl and
the existence of a very large singlet-triplet energy gap in the 
dimerized phase with values $E_g\sim 430K$ for TiOCl \cite{imai_chou} and $~250K$ for TiOBr \cite{powder_tiobr}.

The magnetic Ti chains are arranged in a triangular lattice forming a bi-layer structure. Good Bonner-Fisher fittings to magnetic susceptibility 
measurements are obtained from which it is inferred that these compounds are well described by one-dimensional $S=1/2$ Heisenberg chains.
Moreover, the exchange interaction constants which emerge from those fittings are $J\sim 660K$ \cite{seidel} and $\sim 370K$ \cite{ruckamp} for TiOCl and TiOBr respectively.
These isostructural compounds present two phase transitions: a uniform-incommensurate one at $T_{c2}\sim 92K$ and $\sim 48K$, and an 
incommensurate-dimerized one at $T_{c1}\sim 67K$ and $\sim 27K$ respectively. The order of the high temperature transition has been a subject of extensive controversy
although now there appear to be enough evidence that it is a second order transition \cite{clancy_critical} whereas that at $T_{c1}$ is of first order \cite{seidel}.

We have recently presented a simple microscopic model
of independent Heisenberg chains immersed in a two-dimensional triangular elastic lattice \cite{rpa_ours}. This model predicts 
the appearance of the uniform to incommensurate phase transition through the softening of a phonon near 
the Brillouin zone boundary. This points to the relevant role of the lattice geometry as a driving mechanism for the incommensuration.
On the other hand, a precise characterization of the magnetism inside the 
incommensurate phase is still lacking due to the difficulty of obtaining large enough single crystals to perform 
inelastic neutron scattering experiments. Recently, the dynamical structure factor of TiOBr was obtained by
time-of-flight neutron scattering on powder samples \cite{powder_tiobr}. These measurements suggest the closing of the magnetic gap
when going into the incommensurate phase from the dimerized phase. This seems to be in agreement with our theoretical predictions which will
be detailed in the present work.

In the first part of this article we do a review of the mechanism leading to the incommensurate transition. The model of Ref. \onlinecite{rpa_ours} is
extended to account for incommensuration in the direction perpendicular to the magnetic chains as well. 
Within a model of XY spins coupled to adiabatic phonons, we obtain lattice configurations for different temperatures in the incommensurate phase.
In the second part, we aim to characterize the magnetic excitations in this phase. We search for the existence or absence
of spin gap and calculate spin-spin correlations functions with analytic and numerical techniques.

\section{Lattice geometry as a driving force for the incommensuration}

It is well known that a magnetic chain in which the spins are coupled to the lattice is unstable toward
a dimerization of the ions. This is called a spin-Peierls transition. For a real compound to exhibit this
kind of ordering, quasi-one dimensionality is needed because large interchain magnetic interactions would make
the system develop magnetic ordering before going into the singlet ground state. 
This is a possible reason why few SP materials are known.
Before the discovery of this transition in the TiOX family in 2003 \cite{seidel}, 
the only known inorganic SP compound was CuGeO$_3$. This compound presents a magnetoelastic transition from the
uniform phase at high temperatures to a dimerized phase as it is predicted for a single chain. So the question
is the following: what makes TiOX behave differently so as to develop an incommensurate phase in between?

\begin{figure}[ht]
\includegraphics[width=0.4\textwidth]{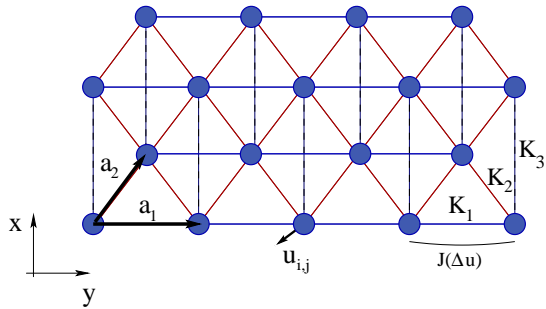}
\caption{\label{figmodel} Schematic representation of our simplified
model. Only Ti atoms are included over the $xy$ plane. $K_1$,
$K_2$ and $K_3$ are the harmonic force constants acting when
two atoms in the same chain, in  neighboring chains
and in next-nearest neighbor chains respectively move from their
equilibrium position. $\textbf{a}_1$ and $\textbf{a}_2$ are the base
vectors. The coordinate of an atom is given by $\textbf{R}_{ij}= i
\textbf{a}_1 + j \textbf{a}_2$.}
\end{figure}

Let us consider the spring-like lattice model of Fig. \ref{figmodel} as representative of the Ti ions in the bilayer structure of TiOX.

\begin{figure}[ht]
\includegraphics[width=0.4\textwidth]{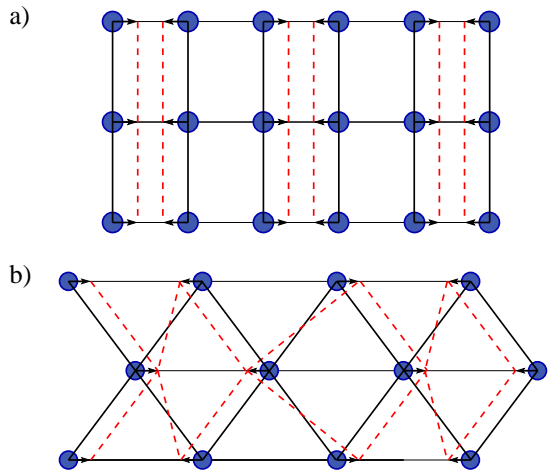}
\caption{\label{dimeriz_sq_tr} Illustrative representation of the action of a dimerization on the
interchain springs in the cases of a) square and b) triangular geometries. Dashed lines represent the interchain spring configurations after the distortion
takes place.}
\end{figure}

We are considering that there are magnetic chains in the $y$ direction which are only elastically coupled between them. $u_{i,j}$ 
represents the displacement of the i-th ion on the j-th chain from its equilibrium
position and $K_1$, $K_2$ and $K_3$ are elastic constants which act longitudinally. 
For simplicity, suppose that the ions move only in the direction of the magnetic chains.
Fig. \ref{dimeriz_sq_tr} clearly shows a marked difference between the situation where the ions in neighboring chains are 
aligned normally and that in which they are shifted by half lattice constant.

If the ions dimerize in an in-phase pattern in the square lattice as shown in Fig. \ref{dimeriz_sq_tr}a, there is no cost of elastic interchain 
energy because the distance to the corresponding ion in the next chain remains unchanged. Therefore, the system will prefer this kind of distortion 
to lower its overall free energy.
This is what happens in the case of CuGeO$_3$. 
On the other hand, for the triangular geometry (Fig. \ref{dimeriz_sq_tr}b) one sees that there is always an increase of interchain energy due to 
the alternating contraction and expansion of the interchain ionic distances. Thus, it is less favorable for the triangular system to dimerize as it 
has to pay more elastic energy than in the square case.

Let us perform a more precise characterization of what would happen with the distortions in the triangular case. Now we are going to allow for movements 
also in the $x$ direction (perpendicular to the magnetic chains) for more generality.
The total mean energy of the elastic system can be calculated as
$E_{ph} = \frac{1}{2}\sum_\textbf{q}\Omega^2(\textbf{q})|Q(\textbf{q})|^2$
where $\Omega(\textbf{q})$ is the phonon dispersion relation and $Q(\textbf{q})$ are the normal coordinates.
We want to study what happens with the elastic energy as we move away from a in-phase dimerization, i.e. from the component $\textbf{q}=(0,\pi)$. 
The square of the lowest dispersion curve \footnote{Note that there are two acoustic dispersion branches within this model as we are working in two 
dimensions. We consider the lowest as the one which couples to magnetism more relevantly.} for our model of Fig. \ref{figmodel} (see
the phononic Hamiltonian $H_{ph}$ in Eq. \ref{hspinph}) reads

\begin{eqnarray}
\label{Wq}
& & \Omega^2(\textbf{q}) = \Omega_1^2 \left(1-\cos{q_y}\right) + 2 \Omega_2^2 \left(1-\cos{\frac{q_x}{2}} \cos{\frac{q_y}{2}}\right)\nonumber\\
& &+ \Omega_3^2 (1-\cos{q_x}) \nonumber\\
& &- \sqrt{\left[\Omega_1^2 (\cos{q_y}-1) + \Omega_3^2 (1-\cos{q_x})\right]^2
+ 4 \Omega_2^4 \sin{\frac{q_x}{2}} \sin{\frac{q_y}{2}}},
\end{eqnarray}

where $\Omega_\alpha^2 \equiv 4K_\alpha/M$.
If we develop this relation near $\textbf{q}=(0,\pi)$ we get
\begin{eqnarray}
%\label{Wqapp}
\Omega^2(\textbf{q}) \sim \Omega_2^2 \left[\left(\frac{q_x^2}{8}-1\right)\delta  + \left(2  + \frac{\Omega_3^2}{\Omega_2^2} -\frac{\Omega_2^2}{4 \Omega_1^2} \right)q_x^2 \right],
\end{eqnarray}
where $\delta\equiv\pi-q_y$.

Therefore we see that the development has a linear term in $\delta$ (or $q_y$) in contrast to what happens for the square lattice in which
the first non vanishing $q_y$ term is of second order. To first order, the square lattice does not alter its elastic energy
in the vicinity of $\delta=0$ but there is an energy gain for the triangular lattice as long as the condition
$q_x<2\sqrt{2}$ is fulfilled. 
This is not a very restrictive condition as we are looking for solutions with $q_x \sim 0$, i.e. a smooth interchain modulation.
As $q_y$ is the component which couples to magnetism in our model, then if the magnetic free energy does not grow very rapidly 
with $\delta$ we expect the total free energy to have a minimum shifted from $\delta=0$. On the other hand, the first non vanishing $q_x$ term
appears at second order. Anyway, for a fixed $\delta >0$, the system would gain elastic energy moving form $q_x=0$ if the condition 
$\Omega_3^2 < \frac{\Omega_2^2}{8} \left(\frac{2 \Omega_2^2}{\Omega_1^2} - \delta \right)$ is satisfied. In conclusion, this model would enable us 
to find an incommensuration in both $q$-directions as seen experimentally. 

\begin{figure}[ht]
\includegraphics[width=0.4\textwidth]{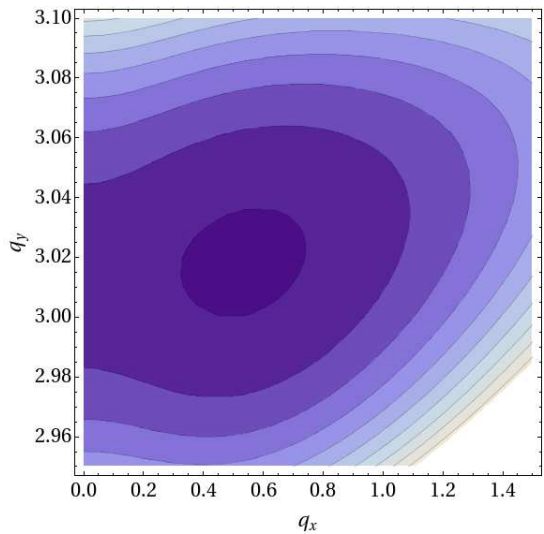}
\caption{\label{trans} Contour plot of the frequency surface of the renormalized phonons at the transition temperature in the area around
$\textbf{q}=(0,\pi)$. The darkest zone encloses the wave vector which acquires the lattice
modulation at the transition point.}
\end{figure}

For a quantitative characterization of the uniform-incommensurate transition, let us extend our model of Ref. \onlinecite{rpa_ours} 
allowing ion displacements in the $x$ direction as well.
We introduce the following Hamiltonian,

\begin{align}
\label{hspinph}
&H=H_{ph}+H_{s}+H_{sph},\\\nonumber
&H_{ph}=\sum_{i,j}{\frac{P_{i,j}^2}{2m}} + \\\nonumber
& \sum_{i,j}\left[\frac{K_1}{2} (u^y_{i,j}-u^y_{i+1,j})^2
+ \frac{K_3}{2}(u^x_{i,j}-u^x_{i-1,j+2})^2\right]+\\\nonumber
&\sum_{\substack{i,j\\ \alpha =\{x,y\}}}\frac{K_2}{4}[(u^\alpha _{i,j}-u^\alpha_{i,j+1})^2 
+(u^\alpha _{i,j}-u^\alpha _{i-1,j+1})^2 +\\\nonumber
& (u^\alpha_{i,j}-u^\alpha_{i,j+1})(u^{\bar{\alpha}}_{i,j}-u^{\bar{\alpha}}_{i,j+1})- \\
&(u^\alpha _{i,j}-u^\alpha _{i-1,j+1})(u^{\bar{\alpha}}_{i,j}-u^{\bar{\alpha}}_{i-1,j+1})],\\
&H_s=J \sum_{i,j}  \textbf{S}_{i,j} \cdot \textbf{S}_{i+1,j},\\
&H_{sph}=\sum_{i,j} \alpha (u^y_{i+1,j}-u^y_{i,j})\,
\textbf{S}_{i,j} \cdot \textbf{S}_{i+1,j}.\\\nonumber
\end{align}

where $P_{i,j}$ is the momentum of the i-th ion on the j-th chain, $K_1$, $K_2$ and $K_3$ are the elastic constants shown in Fig. \ref{figmodel} and 
$\alpha$ is the spin-phonon coupling constant.
The spin-phonon part has been linearized, so the magnetism couples only to the y component of the ion movements to this order. 
It is useful to introduce a dimensionless spin-phonon constant through
\begin{equation}
\lambda =\frac{4\alpha^2J}{\pi M \Omega^2(0,\pi)}.
\end{equation}
The phononic part of this Hamiltonian allowed us to calculate the dispersion relation (\ref{Wq}).
Following the steps of Ref. \onlinecite{rpa_ours}, we calculate the phononic retarded Green's function. 
The phononic self-energy is treated by a \textit{Random phase approximation} approach and we use the expression
obtained in Ref. \onlinecite{cross_fisher} for the dimer-dimer correlation function. There will be a
structural transition when $\mathcal{D}^{ret}(\textbf{q},\omega)$ has a pole with $\omega=0$. This gives us an equation for
the transition temperature as a function of $\textbf{q}$. The $\textbf{q}$ which has the highest temperature will be the wave vector associated to the
lattice distortion at the transition point. Fig. \ref{trans} shows a contour map of the frequencies of the softened phononic dispersion surface at the 
transition temperature, in which it is seen that the  mode which drives the structural transition
is incommensurate for both q-components, in agreement with the experimental measurements \cite{abel, schonleber, smaalen, sasaki, krimmel}.

The free parameters $K_1,K_2,K_3,\lambda$ of out model have been adjusted in order to fit different experimental data
measured in TiOCl (cf. Fig. 3 of Ref. \onlinecite{rpa_ours}): the phonon frequencies obtained by inelastic x-ray scattering for the $(0,q_y)$ 
direction at $T=300K$ \cite{abel},
the value of the bare frequency of the  phonon identified as the one which drives the SP transition \cite{abel}, $\Omega_0 \sim 27 meV$, and $(q_x,q_y)$ values
similar to those found experimentally \cite{abel, schonleber, krimmel}.
Specifically, the values which gave us the best fit were: $\Omega_1=\sqrt{4K_1/M}\sim 6\, meV$, $K_2/K_1\sim 6.7$, $K_3/K_1\sim 11$ and $\lambda=1.3$. 
It is worth saying that this approach overestimates the transition temperature by a factor  $\sim 2.5$ as discussed in Ref. \onlinecite{rpa_ours}. 
We treated this problem in 
Ref. \onlinecite{nonad} showing that an approach using non-adiabatic phonons solves the problem, but we will not do the calculations
for the present dispersion relation since it does not lead to qualitatively different results.

\begin{figure}[ht]
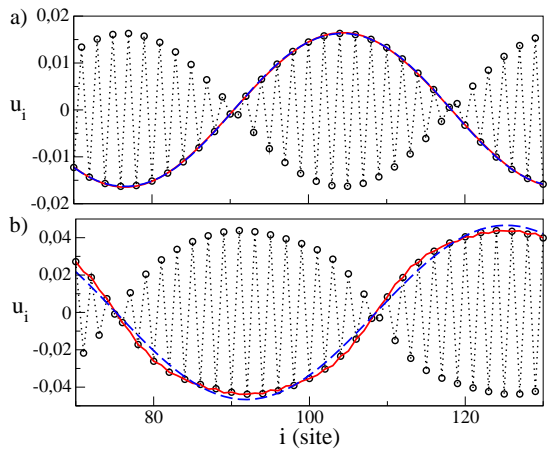

\includegraphics[width=0.4\textwidth]{dist0.14.eps}
\includegraphics[width=0.4\textwidth]{dist0.121.eps}
\caption{\label{dist_incomm} Distortion patterns in the incommensurate phase, a) close to the uniform phase at $T/J \sim 0.14$ 
and b) near the dimerized phase at $T/J \sim 0.12$. Solid lines represent the waves that modulate the distortions given by $(-1)^i u_i$. 
Dashed lines are sinusoidal fits to the envelopes.}
\end{figure}

Now let us turn to follow the evolution of the distortions for different temperatures inside the incommensurate phase. 
For this purpose we propose as a model an array of XY spin chains immersed on the elastic structure of Fig. \ref{figmodel}.
As the XY spin-phonon chain contains the relevant physics of a SP system, we expect that the qualitative 
behavior will be captured by this simplified model, which incidentally can be solved exactly. 
The Hamiltonian is the one defined in (\ref{hspinph}) but now we neglect the z-component of the spin operators.
Now we aim to minimize the total free energy (magnetic + elastic) self-consistently in the displacements at different temperatures. For this
purpose we first perform a Jordan-Wigner transformation on the spin operators of this XY Hamiltonian, which leads to a tight-binding-like model 
of spinless fermions coupled to the lattice. This model can be solved exactly by diagonalizing $N_c$ matrices of size $N_s \times N_s$ (where $N_s$, $N_c$ are the number
of sites and chains) each one, and whose elements depend on the actual displacement configuration. This allows us to obtain the full
one-particle spectrum. Then, the free energy per site is obtained by the expression

\begin{eqnarray}
f = e_{ph} - \frac{T}{N_s N_c} \sum_{l=1}^{N_{c}}\sum_{m=1}^{N_{s}}\ln(1+e^{-\lambda^m_l/T})
\end{eqnarray}

where $e_{ph}$ is the elastic energy density associated with $H_{ph}$ in Eq. \ref{hspinph} and $\lambda^m_l$ is the m-th eigenvalue 
of the matrix associated to the l-th chain.
By this minimization procedure we are able to obtain both phase transitions, i.e. we can account for the full intermediate
incommensurate phase. Fig. \ref{dist_incomm} shows two examples of distortions of one of the chains in the incommensurate phase, 
one near the dimerized phase and the other near the uniform one. Superimposed to each pattern, we show
the wave which modulates the alternation $(-1)^i u_i$, together with a sinusoidal fit to it.

\begin{figure}[ht]
\includegraphics[width=0.4\textwidth]{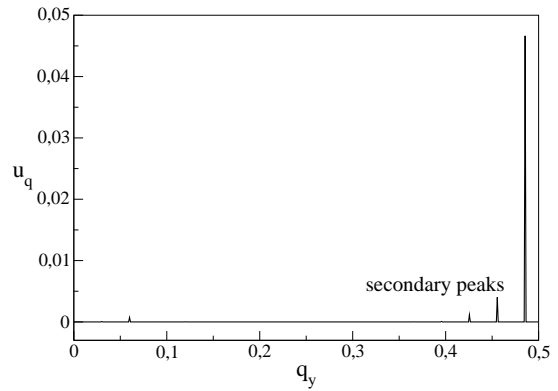}
\caption{\label{secondarypeaks} Fourier transform of the positions shown in Fig. \ref{dist_incomm}b. 
The appearance of secondary peak indicate the tendency to form a soliton lattice near the dimerized phase.}
\end{figure}

We observe that in the case close to the dimerized phase, the modulation is not so well described by
a sine function. The distortion pattern looks more like a soliton lattice which is confirmed by
the secondary peaks appearing in the Fourier transform of the positions (see Fig. \ref{secondarypeaks}). 

When the temperature raises, 
the modulation evolves and close to the uniform phase it is a perfect sine function (Fig. \ref{dist_incomm}a). 
This behavior could explain the experimental temperature evolution of the x-ray peaks observed in Ref. \onlinecite{abel} 
for the incommensurate phase of TiOCl. 
There, a third harmonic of the main incommensurate peak was observed and interpreted as due to a soliton lattice structure. 
Upon warming toward the uniform phase this third harmonic signal reduces and the pattern becomes sinusoidal. 
On the other hand, this behavior is also similar to what it is observed in CuGeO$_3$ when a magnetic field is applied to the compound in the 
low temperature phase \cite{loa}.
In that system, when the magnetic field is high enough to overcome the spin gap (first critical field), the system first develops a soliton lattice and 
with increasing field it tends to a sinusoidal pattern up to a second critical field, above which it becomes uniform.

\begin{figure}[ht]
\includegraphics[width=0.4\textwidth]{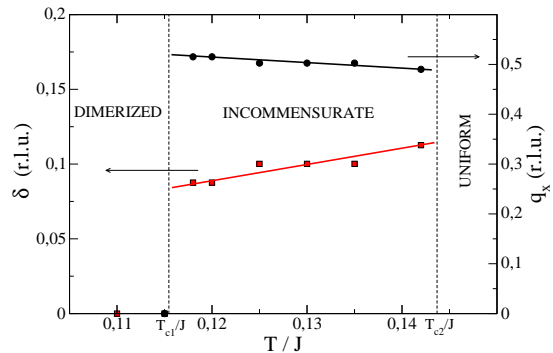}
\caption{\label{q_vs_T} Evolution of $\delta$ and $q_x$ with temperature. Dashed lines are the frontiers which separate the different phases.}
\end{figure}

To finish this section, in Fig. \ref{q_vs_T} we show the evolution with temperature of the wave vector of the main Fourier peak. 

\begin{figure}[ht]
\includegraphics[width=0.4\textwidth]{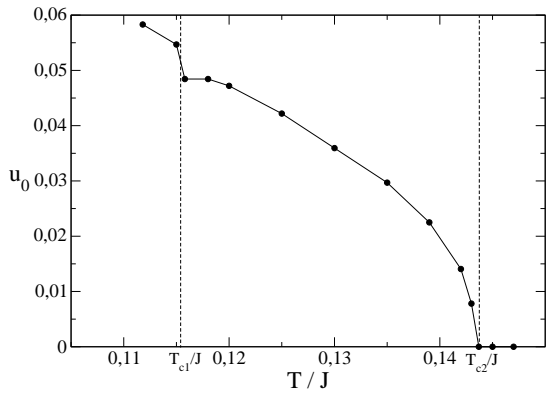}
\caption{\label{u0_vs_T} Evolution of the distortion amplitude with temperature. Note the finite jump of
$u_0$ at $T_{c1}$ and the second order-like behavior around $T_{c2}$.}
\end{figure}

We were able to obtain a similar dependency of $q_y$ with $T$ as seen experimentally \cite{abel,schonleber}.
On the other hand, the $q_x$ modulation that we obtain increases slightly with decreasing temperature, contrary to what it is found experimentally.
This might be due to the fact that our model does not consider magnetic interchain interactions, so the system is not able to gain magnetic energy 
by a distortion in the $x$ direction, in contrast to what occurs
in the direction of the magnetic chains. Fig. \ref{u0_vs_T} shows the evolution of the amplitude of the distortion $u_0$, 
in which we can observe that above $T_{c2}$ we have $u_0=0$ indicating that this is the order parameter for the uniform-incommensurate transition. 

The behavior of this curve around $T_{c2}$ resembles that of a second order phase transition (cf. with Ref. \onlinecite{clancy_critical}). 
On the other hand, at $T_{c1}$ there is a finite jump of
$u_0$ and $\textbf{q}$ (Fig. \ref{q_vs_T}), signaling a first order transition. The order parameter for this
transition is $\delta$.

\section{Nature of the excitations and spin correlations}

Let us now characterize the behavior of the magnetic excitations inside the incommensurate phase. We are going to check for the 
existence or not of a spin gap and calculate spin-spin correlation functions. 
For this purpose we tackle the problem with two different approaches.

\begin{figure}[ht]
\includegraphics[width=0.4\textwidth]{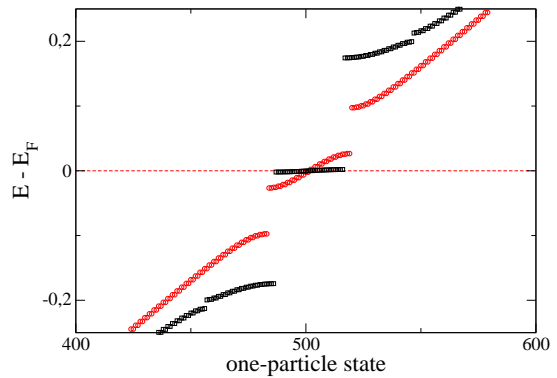}
\caption{\label{spectrum} One-particle energy spectrum for the pseudofermion representation of the XY model with two different modulations. Both cases present continuous bands at the Fermi level, indicating gapless excitations.}
\end{figure}

In the previous section we have minimized the free energy for a system of elastically coupled XY chains and obtained
the corresponding distortions for different temperatures. Now we are going to use these configurations as stable patterns
for one chain and study some properties on it. First we will check the energy spectrum around the Fermi energy of the pseudo-fermion 
chains with hoppings modulated by these distortions. As different $S_z$ subspaces in the magnetic problem are related to a different
filling of the one-particle energy levels in the spinless fermions, this will enable us to identify the presence or absence of a
spin gap. In Fig. \ref{spectrum} we see the spectra for two different patterns. In both cases continuous bands appear at the Fermi level, indicating 
the absence of a gap. We were able to check this behavior for several different patterns which always led to gapless spin excitations.

\begin{figure}[ht]
\includegraphics[width=0.4\textwidth]{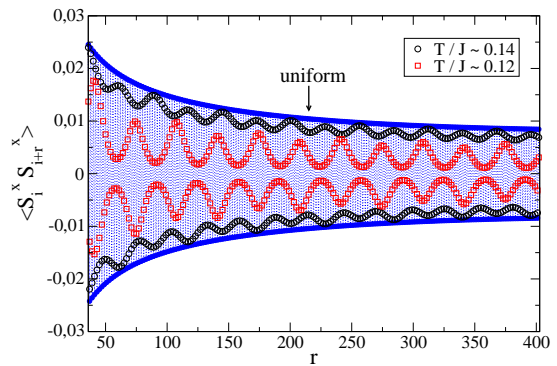}
\caption{\label{corr_xy} Real space spin-spin correlation functions obtained with the XY model for the uniform chain and the 
distortions shown in Fig. \ref{dist_incomm}.}
\end{figure}

\begin{figure}[ht]
\includegraphics[width=0.4\textwidth]{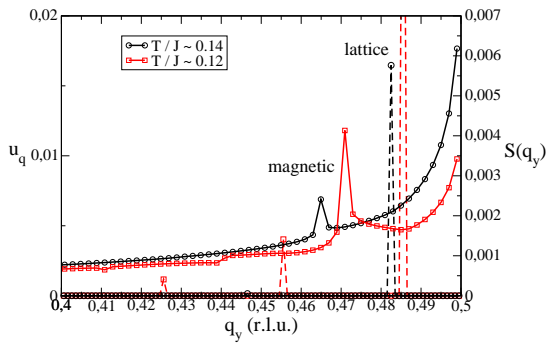}
\caption{\label{fourier_corr_xy} Fourier transform of the correlations shown in Fig. \ref{corr_xy} (solid lines) along with the peaks of the lattice modulation (dashed lines).}
\end{figure}

Now we turn to the calculation of spin-spin correlations $\langle S_i^x S_{i+r}^x \rangle = \langle S_i^y S_{i+r}^y\rangle$. 
This is accomplished through the calculation of $r \times r$ determinants whose elements are obtained by the formula \cite{lsm}

\begin{eqnarray}
G_{i,j}= \sum\limits_{m=1}^{N/2} \left( e_i^m e_j^m - e_i^{m+N/2} e_j^{m+N/2}\right)
\end{eqnarray}

where $e_i^m$ is the i-th component of the eigenvector corresponding to the m-th eigenvalue of the matrix introduced in the previous section.
Fig. \ref{corr_xy} shows real space correlations and Fig. \ref{fourier_corr_xy} shows the static structure factor along with the Fourier transform
of the corresponding distortion pattern. 


On one hand we see that the spin correlations in the incommensurate phase are modulated.
This is made clear by the appearance of new peaks in the structure factor. In every case there appear peaks at $q_y=\pi-2\delta$ and $q_y=\delta$
(not shown in Fig. \ref{fourier_corr_xy}) besides the one at $q_y=\pi$ which already belongs to the uniform antiferromagnetic lattice.
With this information in hand we were able to fit the real space correlations by the following formula

\begin{eqnarray}
\label{sscorrelxy}
\langle S_i^x S_{i+r}^x\rangle = a_0 \frac{\cos{\delta r}}{\sqrt{r}} + a_1 \frac{(-1)^r (1+a_2 \cos{2\delta r})}{\sqrt{r}}
\end{eqnarray}

where the parameters $a_i$, which measure the relative intensity of each peak, take on different values for each distortion.
On the other hand, the magnetic Fourier peak near $\pi$ is at twice the distance from $\pi$ of the lattice peak. It is worth mentioning
that the peak at $q_y=\delta$ has a much smaller intensity than those at $q_y=\pi$ or $q_y=\pi-2\delta$.

\begin{figure}[ht]
\includegraphics[width=0.4\textwidth]{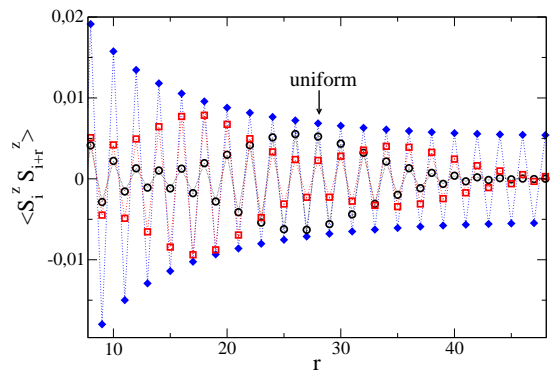}
\caption{\label{corr_dmrg} Real space spin-spin correlation functions obtained with DMRG for two different single-chain configurations (signaled
with squares and circles) and the result for the uniform chain as a reference.}
\end{figure}

\begin{figure}[ht]
\includegraphics[width=0.4\textwidth]{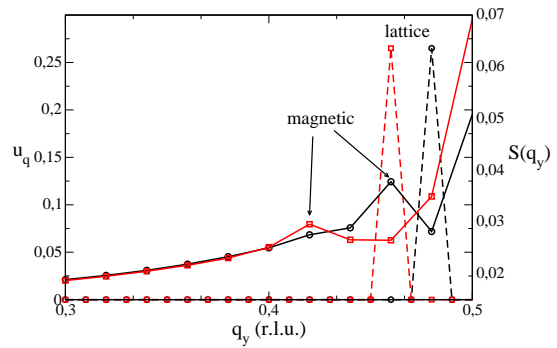}
\caption{\label{fourier_corr_dmrg} Fourier transform of the magnetic correlations of Fig. \ref{corr_dmrg} (solid lines) along with the peaks of the lattice modulation
(dashed lines). The symbols are in accordance with those of that figure.}
\end{figure}

We now turn to the study of the Heisenberg model with modulated exchange on a single chain. Nowadays, the Density 
Matrix Renormalization Group (DMRG) numerical technique appears as the ineluctable numerical tool for treating static 
properties in one dimensional interacting systems at zero temperature \cite{schollwock,hallberg}. For our purpose 
we chose its finite size algorithm with periodic boundary conditions (PBC) on systems of sizes up to $L=200$, instead of open 
boundary conditions to avoid edge effects. Even though it is known that using PBC leads to a slower convergence of the method \cite{white}, 
keeping just 300 states was enough to get quasi-exact 
results with a discarded weight of order $O(10^{-12})$. In particular we are interested in the spin gap, here determined as 
$\Delta_s=E_0(S_z=1)-E_0(S_z=0)$, being $E_0(S_z)$ the ground state energy on a chain in the subspace of a given total $S_z$. 
Different single-chain distortion patterns of the type $u_i=u_0 \cos (\textbf{q} \cdot \textbf{R}_i)$ were tried and 
the absence of spin gap was found for every configuration, confirming a spin gapless situation in the 
incommensurate phase. Spin-spin correlation functions have been also calculated and the 
results are shown in Figs. \ref{corr_dmrg} and \ref{fourier_corr_dmrg}.

As for the case of the XY model, the real space correlations show a power law behavior, indicative of the absence of spin gap. 
The decay is modulated by an envelope whose Fourier component is $q=\pi-2\delta$ as found previously.
We propose, as a check for this prediction, the measurement of the static magnetic structure factor by elastic neutron scattering. 
This should present an incommensurate peak at twice the distance from the commensurate peak of the low temperature phase with respect to that 
seen in the lattice structure factor.

To conclude this section, let us comment on the results given by bosonization on the Heisenberg Hamiltonian and a Renormalization Group analysis.
This procedure has been treated in Ref. \onlinecite{senlal} for a problem of fermions with incommensuration. Here, we reproduce the bosonized version
of the Hamiltonian using our parameters, which reads
\begin{eqnarray}
\label{hambos}
 H&=&\frac{1}{2\pi} \int dx \left[uK (\pi \Pi (x))^2 + \frac{u}{K} (\nabla\phi(x))^2\right] \nonumber\\
 &+& \frac{\alpha u_0}{2 \pi \bar{\alpha}} \int dx \left[\sin(2\phi + \delta x) + \sin(2\phi - \delta x)\right]
\end{eqnarray}
where $\phi$ is a smooth bosonic field, $\Pi$ is the field conjugated to $\phi$, $u$ is a velocity, $K$ is the Luttinger parameter and $\bar{\alpha}$ is a cutoff of the theory.

The flow equation for $\delta$ is given by $\frac{\textrm{d} \delta}{\textrm{d} l}=\delta$ (cf. Eq. 45 of Ref. \onlinecite{senlal}). 
It is easy to observe that $\delta$ grows with scale, which means that the sine functions in (\ref{hambos}) oscillate rapidly so the integral will vanish.
Therefore we end up with a free bosonic theory indicating that the excitations will be gapless as in a uniform Heisenberg chain.
Note that the bosonized Hamiltonian of Eq. (\ref{hambos}) could correspond to a general XXZ model. 
The only change is the value of the Luttinger parameter $K$ \cite{giamarchi}. 
Therefore, the results presented in this Section seem to indicate that the behavior of the correlation function given in Eq. (\ref{sscorrelxy}) 
has an universal form which is obtained by changing the exponent of the power law decay from a square root in the XX model, 
to a K-dependent exponent in a general anisotropic XXZ model.

\section{Conclusions}
In conclusion, we have studied the behavior of a system of spin-$1/2$ magnetic chains which are immersed in a triangular planar
lattice. It is seen that this triangular geometry leads the system to a quite different behavior compared to a regular spin-Peierls
mechanism, like that found in CuGeO$_3$ and other organic compounds. The elastic frustration produced by the triangular arrangement of the magnetic ions
leads the system to produce an incommensurate phase between the uniform and dimerized ones. This is what it is observed experimentally
in the compound family TiOX (X = Cl, Br). We have seen that with a simple model of magnetically uncoupled chains we were able to obtain
both phase transitions. Moreover, by a self-consistent minimization of the free energy, we could describe the lattice configurations
at different temperatures which allowed us to follow the y-component of the main modulation wave-vector with temperature.
On the other hand, spin-spin correlation functions were calculated exactly with XY chains and by DMRG calculations on the Heisenberg model
on single chains with sinusoidal modulations. In both approaches we obtained power law behaviors at long distances, modulated by an envelope
which has a Fourier peak at $2\delta$, where at $\delta$ is the peak of the imposed lattice modulation. Moreover, this power law decay of
the correlations is indicative of the absence of spin gap in the excitation spectrum, which was confirmed by looking at the one-particle
spectrum in the XY case, and the vanishing energy difference between $Sz=0$ and $Sz=1$ subspaces in the case of the DMRG calculations.
Furthermore, an analysis of the bosonized version of the spin-phonon Hamiltonian together with the Renormalization Group flow equation
of the incommensuration parameter $\delta$ led to the same conclusion. 
In recent time-of-flight neutron scattering experiments in TiOBr, it has been shown that the spectral weight increases at low energy when 
going from the dimerized to the incommensurate phase \cite{powder_tiobr}.
This is in agreement with the theoretical prediction of the present paper that the gap closes in the incommensurate phase.
A more detailed experimental determination of the magnetic spectrum and the theoretical calculation of the dynamical correlation 
function in the incommensurate phase would be needed to further clarify the nature of the magnetic excitations in this uncommon phase 
which we have characterized as an incommensurate spin Luttinger liquid phase.

\begin{acknowledgments}
The authors would like to acknowledge J. P. Clancy for helpful discussions.
This work was partially supported by PIP 11220090100392 of CONICET, PICT 1647 and PICT R 1776 of ANPCyT.
\end{acknowledgments}

\bibliography{refs}
\bibliographystyle{apsrev4-1}

\end{document}